# About the annual distribution of magnetically disturbed cloudless days and nights over Abastumani (41.75°N, 42.82°E)

**G.G. Didebulidze***, **G.Sh. Javakhishvili, M.A. Marsagishvili, M. Todua**

Georgian National Astrophysical Observatory, Ilia Chavchavadze State University, A. Kazbegi ave. 2a, 0160 Tbilisi, Georgia

**Abstract**

The annual distribution of the planetary geomagnetic *Ap* index monthly mean values on visually cloudless days and nights at the Abastumani Astrophysical Observatory (41.75°N, 42.82°E) in 1957-1993 is analyzed. Cloudless days were selected from day-time observations of the total ozone content and cloudless nights from night-time measurements of the mesosphere-thermosphere nightglow intensity. Both series of observations were carried out almost uninterrupted during the referred time period. The distribution of long-term monthly mean *Ap* values on cloudless days shows a semi-annual variation with maxima on the equinoctial months (March and September), while on cloudless nights it displays a seasonal-like character. The semi-annual and seasonal-like variations of *Ap* modulate the annual distribution of long-term monthly numbers of cloudless days with maximum in August and that of cloudless nights with maximum in September. The lowest monthly mean *Ap* value for cloudless days and its highest value for cloudless nights fall on June. Different annual distributions of long-term monthly numbers of cloudless days and nights, when sudden storm commencements are present, have been found. This coupling between the visually cloudless days/nights and the magnetically disturbed days may be considered as a manifestation of the influence of space weather/cosmic factors on the regional and global climate.

* Corresponding author. Fax: (995 32) 376303, E-mail: didebulidze@genao.org

# 1. Introduction

The problem of the impact of different cosmic factors on cloud cover and global climate has been of increasing interest during the last decade. For understanding the global warming processes and their consequences, it is crucial to distinguish these effects from other natural and anthropogenic factors affecting cloud cover, to model their influence on the circulation in the lower and upper atmosphere and on the atmosphere/ionosphere coupling processes (Rishbeth and Roble, 1992).

Galactic cosmic rays (GCR) are thought to be one of the major cosmic factors affecting cloud covering (Svensmark and Friis-Christensen, 1997). The total cloud cover and the GCR flux modulated by solar activity are correlated, which may reflect the connection between solar and climate variability involving the solar wind, GCRs and clouds (Marsh and Svensmark, 2000). The solar wind carrying solar plasma by the interplanetary magnetic field modulates the flux of GCRs, which are the main source of ionization in the troposphere/ lower stratosphere and can initiate the formation of cloud condensation nuclei (Tinsley et al., 2006, and references therein). The solar variations also influence geomagnetic activity via the interplanetary magnetic field. During active solar processes, such as solar flares and coronal mass ejections, the solar wind plasma flux increases in the interplanetary space, which causes a GCR flux decrease - so-called Forbush decrease (Kniveton, 2004; Kudela and Brenkus, 2004) and generates geomagnetic disturbances.

Using 20-year observations, Kudela and Brenkus (2004) found only several Forbush decreases without geomagnetic storms and very rare cases of geomagnetic storms without the Forbush decreases. The Forbush decrease may possibly result in a reduction of cloud cover amount (Kniveton, 2004; Svensmark and Friis-Christensen, 1997; Marsh and Svensmark, 2000). The solar wind-magnetosphere interactions are efficient when the southward component of the interplanetary magnetic field is parallel to the geomagnetic field (magnetic dipole). Such favourable conditions for development of geomagnetic storms mainly occur at the equinoxes (Chaman-Lal., 1998; Russell and McPherron, 1974). These phenomena are thought to be a source of the well-known semi-annual variations in the magnetically disturbed day number with maxima at vernal and autumnal equinoxes and minimum at the summer solstice.

The annual distribution of cloud cover on the globe and in separate regions should reflect its seasonal peculiarities, as well as its probable coupling with variations in physico-chemical processes in the atmosphere accompanying geomagnetic disturbances, which depend on the Sun-Earth geometry.

It is known that, in some regions on the globe, the cloud cover variations correlate with the solar cycle and not GCRs (Udelhofen and Cess, 2001). In this case the cloud variability may be affected by the modulation of the atmospheric circulation resulting from the variations in the solar-UV-ozone-induced heating (Udelhofen and Cess, 2001). Hence, the mean annual distribution of cloudless days and nights covering several solar cycles could help us to distinguish the space weather/cosmic factors affecting the regional and global climate.

In the present paper, we consider the annual distribution of the planetary geomagnetic *Ap* index values for visually cloudless days and cloudless nights (hereafter denoted as CD and CN, respectively) at the Abastumani Astrophysical Observatory (41.75°N, 42.82°E) in 1957-1993. The planetary geomagnetic *Ap* index (which characterizes the coupling between the mid-latitude lower and upper atmosphere) is used for describing magnetically disturbed days and nights. We also analyze the annual distribution of CD and CN when a sudden storm commencement (SSC) occurs.

## 2. The annual distribution of magnetically disturbed cloudless days and nights

The annual distribution of monthly numbers of cloudless days and nights, in addition to other meteorological parameters, is one of the major characteristics of the global and regional climate. This distribution depends on geographical region and it is one of the important features of the climate of an observatory. At the Abastumani Astrophysical Observatory experimental investigations of the Earth's atmosphere by optical methods are being carried out on visually cloudless days and cloudless nights (Megrelishvili, 1981; Fishkova, 1983; Didebulidze et al., 2002). The minimum field of view from the observatory corresponds to the radius of the visually low (with altitude up to 3.5 km) cloudless region $\geq 9$ km and the maximum to the radius $\geq 20$ km. Since 1957 within a radius of about 40 km around the observatory, there have been no sources of anthropogenic (industrial etc.) aerosols, which could interfere in the cloud covering processes above the observatory. Therefore, this site is favourable for studying the influence of space weather/cosmic factors on the cloud covering processes. During the last decade these phenomena have been considered to be important for understanding global climate changes (Marsh and Svensmark, 2000; Svensmark and Friis-Christensen, 1997; Udelhofen and Cess, 2001).

In this paper we investigate the annual distribution of the monthly mean geomagnetic planetary $Ap$ index values for visually CD and CN at Abastumani. In 1957-1993, measurements of the total ozone content (TOC) and the mesosphere-thermosphere nightglow intensity had been carried out on most cloudless days (around LT 8h00m-14h00m) and nights, which encompasses three solar cycle interval. The total number of CD is 4323 and CN 1534. Number of days in months vary from 227 to 531 for CD and from 78 to 199 for CN. Note that nightglow observations were carried out during moonless conditions. This allows us to state that the annual distribution of the monthly mean $Ap$ values may be considered as an average annual characteristic of the processes of the space weather-cloud cover coupling.

In Fig.1, the annual distributions of monthly mean values of the planetary geomagnetic $Ap$ index (dashed lines with rectangles) for $Ap<50$, during the period of 1957-93, are plotted. Solid line and circles represent mean $Ap$ for CD (panel A) and CN (panel B). $Ap<50$ correspond to weak and moderate geomagnetic disturbances. We do not consider here strong geomagnetic disturbances which are rarely encountered – in less then 5% cases. Although the mean annual distributions for all $Ap$ are similar to those for $Ap<50$, the dispersions of large $Ap$ values are too big to achieve statistically significant results.

The distribution of $Ap$ has a semi-annual character with a maximum on March for CD (Fig.1A) and exhibits seasonal variations with maxima on February, April, June and September for CN (Fig.1B). The monthly mean $Ap$ values have approximately the same distribution for both CD and CN around September (with $Ap \approx 15$), while during the other seasons they differ. The biggest difference takes place in June, when the mean $Ap$ index value for CD is 11.2 (427 days), while for CN it is 14.4 (111 nights).

Note that a similar annual distribution has been obtained for the sum of the planetary geomagnetic $Kp$ indices ($\Sigma Kp$), which is not shown here for brevity. Both indices are used for study of many dynamical processes in the mid-latitude atmosphere and ionosphere under different geomagnetic conditions. We calculated correlations between two relative monthly numbers: the ratio of magnetically disturbed CD to all CD and the ratio of magnetically disturbed days to all days, in the cases of both indices. The correlation was a little higher using $Ap$ index, therefore we preferred it in this study.

The difference between the annual distributions of the monthly mean $Ap$ index values for CD and CN should be reflected on the annual distributions of numbers of CD and CN. In the following we consider these distributions of monthly numbers of all visually CD and CN, those for all magnetically disturbed day-night periods ($Ap \geq 12$), for relatively strong magnetically disturbed day-night periods ($Ap \geq 20$, $Ap \geq 30$) and when SSCs occurred. This is demonstrated in Fig. 2: the annual distributions of total monthly numbers of visually CD (circles, panel A), CN (dark circles, panel B), magnetically disturbed day-night periods with $Ap \geq 12$ (squares), $Ap \geq 20$ (diamonds) and $Ap \geq 30$ (triangles) and SSCs (stars) at Abastumani in 1957-93 are plotted. The planetary geomagnetic index $Ap \geq 12$ (squares) covers all magnetically disturbed day-night periods. The total numbers of CD and CN when SSCs occured are 367 and 143, respectively.

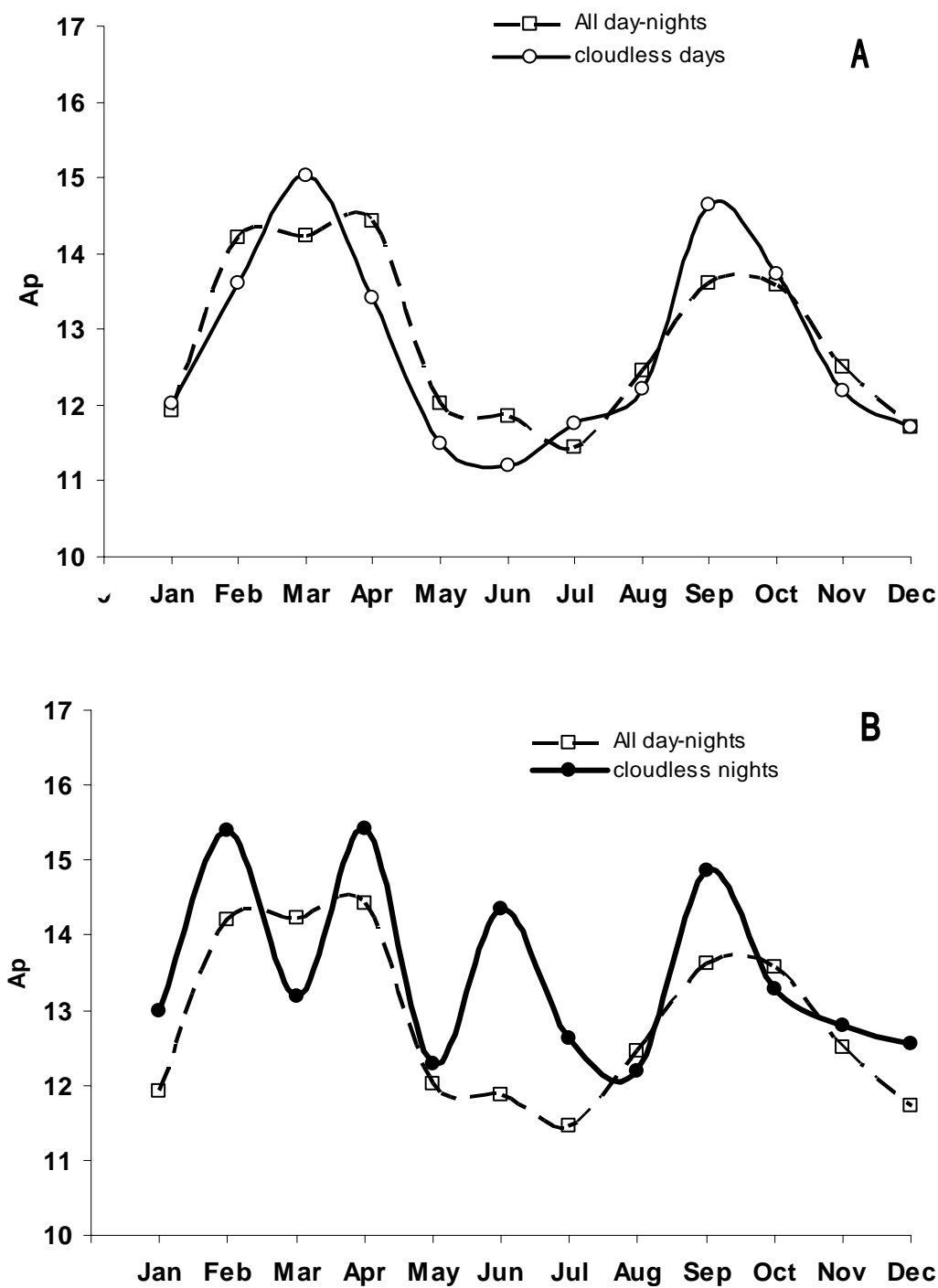

Fig. 1. The annual distributions of monthly mean values of the planetary geomagnetic *Ap* index for *Ap<50* during all day-night periods (dashed line and squares), cloudless days (full line and circles, panel A) and nights (full line and dark circles, panel B) at Abastumani in 1957-1993.

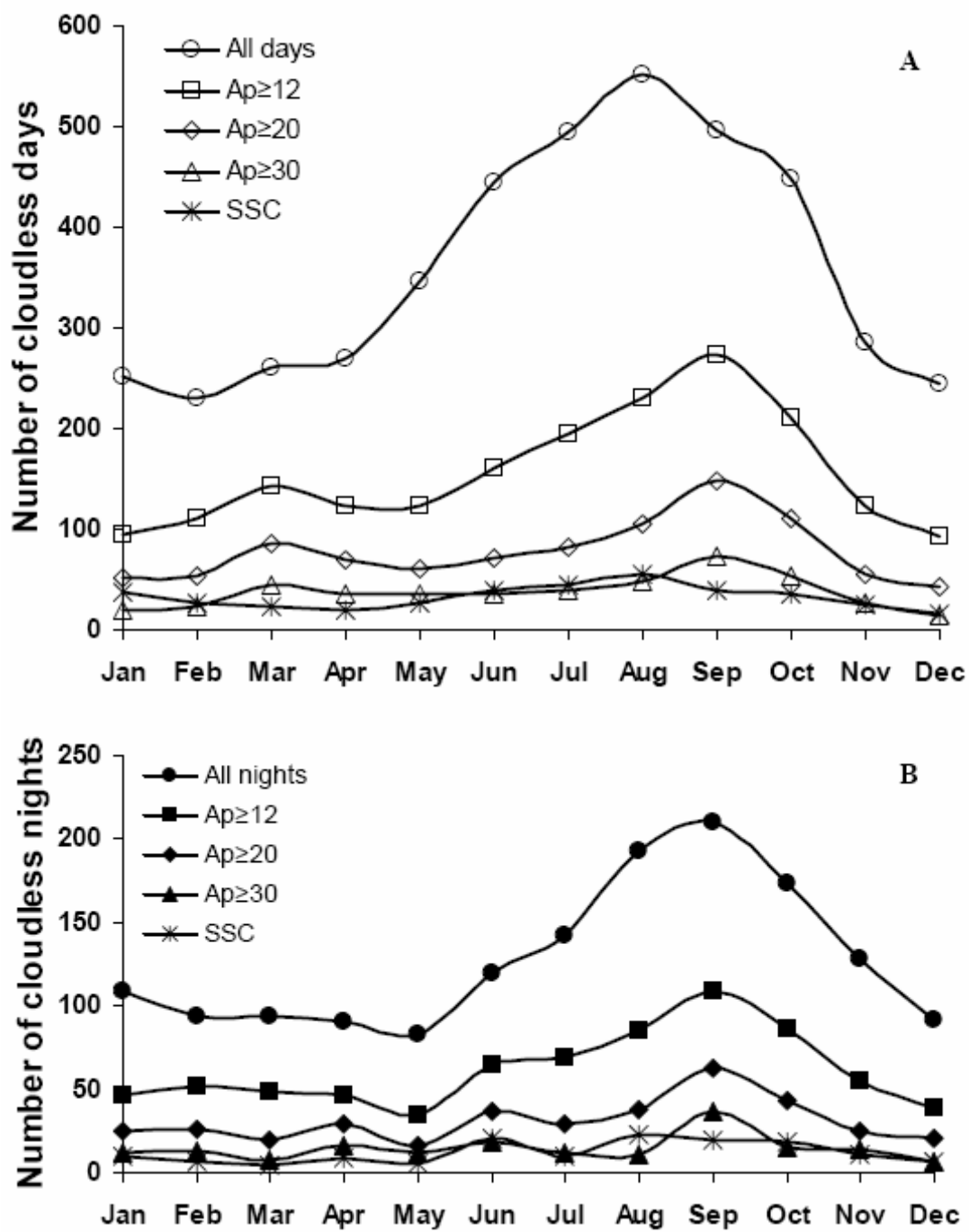

Fig. 2. The annual distributions of total monthly numbers of visual cloudless days (circles, panel A), cloudless nights (dark circles, panel B), magnetically disturbed day-night periods with SSC (stars) and the planetary geomagnetic index $Ap \geq 12$ (squares), $Ap \geq 20$ (diamonds) and $Ap \geq 30$ (triangles) at Abastumani in 1957-93.

Fig. 2 shows that the monthly numbers of CD and CN undergo annual variations with maxima in August (cloudless days) and September (cloudless nights). The annual distributions of those for magnetically disturbed ($Ap \geq 12$, $Ap \geq 20$, $Ap \geq 30$) CD and CN (for short we will use MDCD and MDCN, respectively) are modulated by the semi-annual and seasonal distributions of CD and CN. These modulations are similar to the semi-annual and seasonal distributions of mean monthly values of the $Ap$ index for CD (Fig. 1A) and CN (Fig. 1B).

SSCs occur during the day-night periods with various values of the *Ap* index, at strong, as well as relatively weak geomagnetic disturbances. Total monthly numbers vary from 82 to 98. In this case we additionally consider the annual distributions of monthly relative numbers of SSCs for CD and CN, which we define as a ratio of the monthly numbers of SSCs on CD and CN to the total monthly numbers of CD and CN. These are plotted in Fig. 3, dashed line and squares represent all day-night periods, thin line and circles CD, and dark line and triangles CN. This figure also shows a difference between the annual distributions for MDCD and MDCN. Here a noticeable distinction of the annual distribution of SSC relative numbers for CN from those for CD and for all day-night periods takes place in June again. This relative number for CN is about 0.18 in June and about 0.04 in March. During the day-time and on CD there are no considerable changes in SSC occurrence (their relative numbers are about 0.08-0.1).

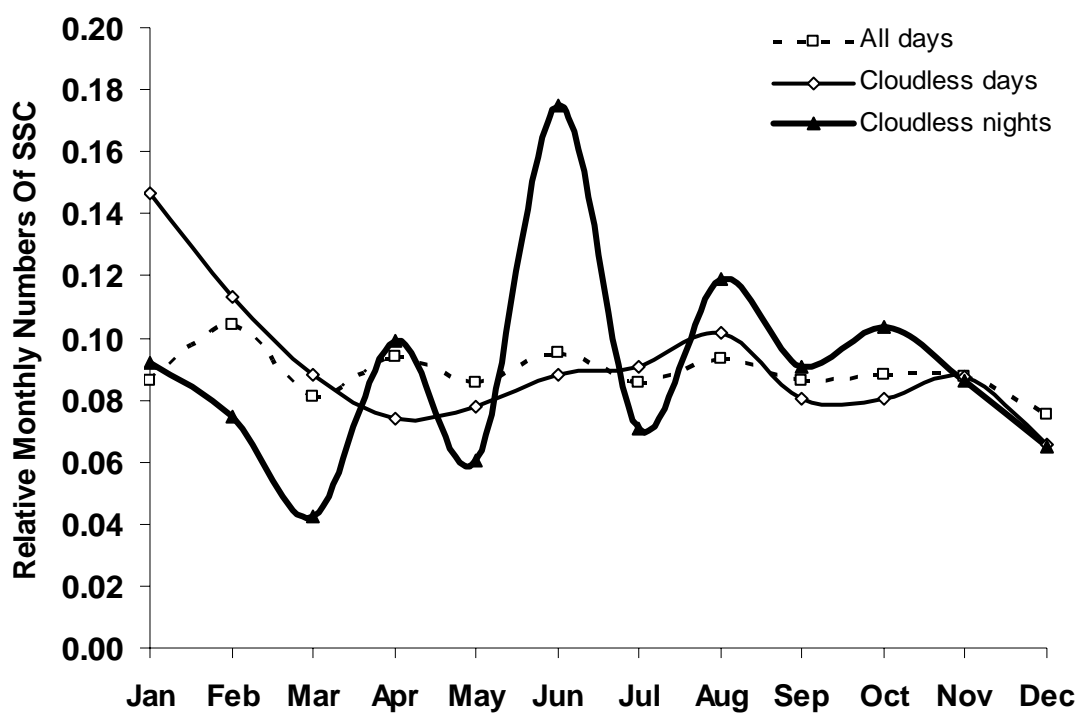

Fig. 3. The annual distributions of relative total monthly numbers of SSC for all day-night periods (dashed line and squares), cloudless days (thin line and circles) and nights (dark line and triangles) at Abastumani in 1957-1993.

### 3. Discussion

A difference between the annual distributions of MDCD and MDCN, found in this study, could be a manifestation of different physico-chemical processes in the troposphere-lower stratosphere and different conditions for cloud formation for day- and night-time. Different sensitivity of cloud covering to cosmic factors during day- and night-time may be caused by a superposition of the troposphere-lower stratosphere physico-chemical processes with the regional and global circulation. The variations in the geomagnetic field do not affect directly the physico-chemical processes in the troposphere/stratosphere. The variations in the atmospheric parameters and geomagnetic activity would be rather a result of their relation to solar activity phenomena, such as the solar proton events (SPEs), the X-ray and UV-radiation from solar flares and the modulation of the GCR flux. Strong changes in the dynamical and structural processes taking place in the lower and upper atmosphere during the geomagnetic events are mostly observed in the polar regions (Jackman et al., 1995; Elsner and Kavlakov, 2001). In magnetically disturbed day-night periods important dynamical variations occur in the atmosphere and ionosphere at polar latitudes (>68º N) during the decay of auroral electrojet. The atmospheric waves and the wind generated during these events can reach the mid-latitude thermosphere-ionosphere F2 region and cause variations in the neutral and charged particle density and the nightglow intensity (Fishkova, 1983; Shefov et al., 2006). But for the mid-latitude lower atmosphere any dynamical and structural changes in magnetically disturbed day-night periods are unknown. Solar protons and relativistic electrons accompanied by geomagnetic storms (*Ap* is mainly >40) can affect the chemical composition of the atmosphere in the polar regions. Only very strong SPEs can influence the chemical composition (the ozone concentration) of the mesosphere-stratosphere at higher (>50º N) middle latitudes (Jackman et al., 1995; Laštovička and Križan, 2005 and references therein). Such events are very rare, and their influence on the mid-latitude upper troposphere and stratosphere is not considered. In this study the annual distribution of *Ap* (Fig.1) modulates the annual distribution of CD and CN, which is noticeable for weak magnetically disturbed days and nights with *Ap>12* for mid-latitude region.

If we assume that GCRs are a major cosmic factor affecting the density of cloud condensation nuclei, then number of magnetically disturbed day-night periods (including SSCs) may probably be accompanied by the Forbush decreases and the reduction of cloud cover. If there were no significant differences in the regional circulation affecting cloud cover during the day- and night-time, then the cloud covering processes under the influence of GCRs or other cosmic factors accompanying magnetic disturbances will be similar for both periods. The difference between the annual distributions of total monthly numbers of CD with maximum in August (Fig. 2A) and CN with maximum in September (Fig. 2B) could result from the different solar ultraviolet radiation absorption by the stratospheric ozone and the related different regional circulation in the troposphere-lower stratosphere. In this case the modulations of the annual distributions of MDCD and MDCN are also different (as noted above). The relatively large monthly mean *Ap* index value on CN in June (Fig. 1B) may be related to a decrease in GCRs, but its minimum value on CD (Fig. 1A) cannot be explained by the same mechanism. The inverse picture in March – a day-time increase in *Ap* accompanied by its night-time decrease - cannot be explained by a decrease in GCRs either. The behaviour of the annual distributions of monthly mean *Ap* is similar during the day- and night-time mainly in autumn. In this case the number of magnetically disturbed cloudless day-nights increases in September, which was expected assuming the influence of GCRs on cloud cover.

Hence, the different annual distributions of the monthly mean values of the geomagnetic *Ap* index for CD and CN and the similar modulations of the annual distributions of total monthly numbers of MDCD and MDCN show the impact of cosmic factors on regional cloud cover. The source of these phenomena cannot be only GCRs, therefore it is necessary to consider some additional factors responsible for variations in cloud cover and regional climate. Such factors would accompany magnetical disturbances. The absorption of solar ultraviolet radiation by the stratospheric ozone during the day-time affects the stratospheric structure as well as regional circulation. Variations in the TOC in magnetically disturbed day-night periods are known (Belinskaya et al., 2001; Laštovička and Križan, 2005, and references therein). Regional peculiarities in the lower and upper cloud covering during the day- and night-time should be regarded as a manifestation of the complex interplay of various terrestrial and cosmic factors. A more detailed consideration of these phenomena needs additional data on the tropospheric aerosol vertical distribution under different helio-geophysical conditions. The major variations in the annual distribution of MDCD and MDCN in the equinoctial and the solstice months (Figs. 1-3) indicate again the role of the Sun-Earth geometry for the mean annual variations in the global and regional climate and their variability under various cosmic factors.

The sensitivity of the cloud cover amount to geomagnetic disturbances should be taken into account when analyzing the middle and upper atmospheric parameters obtained by the ground based and in situ/satellite methods. Such parameters, as the stratospheric TOC (or its height distribution) and the mesosphere-thermosphere nightglow intensity, whose ground based measurements are carried out on cloudless days/nights, are sensitive to magnetic disturbances (Fishkova, 1983; Shefov et al., 2006). In this case, the long-term or annual variations in these middle and upper atmosphere parameters may not be identical.

## 3. Conclusion

We have analyzed the annual distributions of the long-term monthly numbers of cloudless days and cloudless nights and the corresponding monthly mean values of the planetary geomagnetic *Ap* index over the Abastumani Astrophysical Observatory in 1957-1993. For cloudless days, the annual distribution of the monthly mean *Ap (<50)* values shows a semi-annual variation with maxima in the equinoctial months (March and September), while for cloudless nights, it displays a seasonal character. In the summer (June), a minimum mean *Ap* value of about 11.2 for cloudless days and a maximum mean value of about 14.4 for cloudless nights were found.

A modulation of the annual distribution of long-term monthly numbers of cloudless days (with maximum in August) and nights (with maximum in September) by semi-annual and seasonal variations has been revealed for various levels of geomagnetic disturbances (*Ap*≥12, *Ap*≥20 and *Ap*≥30).

Different annual distributions of relative monthly numbers of SSC for CD and CN have also been found. For cloudless nights the largest relative number of SSC appears in June.

These phenomena are considered as a result of the influence of space weather/cosmic factors on the Earth's climate.

**Acknowledgements**
This work has been supported by the ISTC G-1074 and INTAS grant 03-51-6425.